\documentclass[prl,twocolumn,showpacs,preprintnumbers,amsmath,amssymb,citeautoscript]{revtex4-1}
\usepackage{graphicx}% Include figure files
\usepackage{dcolumn}% Align table columns on decimal point
\usepackage{bm}% bold math
\usepackage{color}
\usepackage{textcomp}

\begin{document}
%\preprint{ver3.0}

\title{Zero-field current-induced Hall effect in ferrotoroidic metal}

\author{K.~Ota$^1$}
\author{M.~Shimozawa$^{1,{\dag}}$}
\author{T.~Muroya$^1$}
\author{T.~Miyamoto$^1$}
\author{S.~Hosoi$^1$}
\author{A.~Nakamura$^2$} 
\author{Y.~Homma$^2$}
\author{F.~Honda$^{2,3}$}
\author{D.~Aoki$^2$}
\author{K.~Izawa$^1$}

\affiliation{
$^1$Graduate school of Engineering Science, Osaka University, Toyonaka 560-8531, Japan\\
$^2$Institute for Materials Research, Tohoku University, Oarai, Ibaraki 311-1313, Japan\\
$^3$Central Institute of Radioisotope Science and Safety, Kyushu University, Fukuoka 819-0395, Japan\\
$^{\dag}$\rm{To whom correspondence should be addressed.\\
E-mail: shimozawa@mp.es.osaka-u.ac.jp}
}

\date{\today}

%\pacs{xxx}

\begin{abstract}
We have performed precise Hall measurements for the ferrotoroidic candidate material UNi$\bf{_4}$B. Below N$\acute{\rm e}$el temperature $T_{\rm N}\sim20\ {\rm K}$ (corresponding to the ferrotoroidic transition temperature), a Hall voltage becomes finite even at zero field and changes proportional to the square of current density; by contrast, it is almost zero above $T_{\rm N}$. Moreover, we have found that a current-induced magnetization estimated from our Hall effect measurements is qualitatively consistent with the previous directly measured value. These results provide strong evidence for a magnetoelectric phenomenon uniquely in ferrotoroidic metals---a zero-field nonlinear Hall effect resulting from the current-induced magnetization connecting the ferrotoroidal moments.
\end{abstract}

\maketitle
%%%%%%%%%  Introduction  %%%%%%%%%
The electromagnetic properties of matter often involve breaking the symmetry of the system. Among the simplest examples are two types of elementary electromagnetic dipoles \cite{Jackson62}; an electric dipole arising from a separation of opposite charges requires a lack of spatial-inversion symmetry, and a magnetic dipole due to a circular current needs a broken time-reversal symmetry. By contrast, a third elementary electromagnetic dipole, namely a magnetic toroidal (MT) dipole \cite{Zel'dovich58}, is generated by a vortex of magnetic dipoles including atomic spins or orbital currents, which violates both of the symmetries (see Fig.\ 1(a)). The space- and time-asymmetricity in principle allows cross-coupling between electricity and magnetism in a solid \cite{Curie94,Ederer07,Spaldin08}, and this so-called magnetoelectric effect is of great interest both fundamentally and technologically. Therefore, the magnetoelectric response with respect to the MT dipoles has been intensively studied, just like the case of `insulating' multiferroics with the coexistence of two order parameters \cite{Eerenstein06,Arima11}, but their research was limited mainly to insulators \cite{Sannikov97, Popov98, Popov99, Aken07,Zimmermann14} due to lack of good candidate metals.

Recently, the subject of the magnetoelectric effect has started to move toward metallic materials. According to the theoretical works based on the odd-parity augmented multipole, a certain antiferromagnetism can be regarded as a ferrotoroidic metallic state that accompanies the spontaneous parallel alignment of MT dipoles \cite{Hayami14, Yanase14, Nakatsuji15, Suzuki17, Watanabe17, Hayami18}. In this state, an electric current has been suggested as a control parameter that drives magnetoelectric phenomena \cite{Hayami14}; indeed, several experiments have pointed out that a net magnetization is generated by an electric current flowing in ferrotoroidic ordered states \cite{Saito18, Shinozaki20}. Even more interestingly, the electric current has been proposed to bring about a unique magnetoelectric response for conductive materials, such as a nonlinear Hall effect originating from the current-induced magnetization \cite{Hayami14, Khomskii14}. However, because the amplitude of the applied electric current is limited due to a Joule heating, the magnetoelectric response associated with metals would generally be very small compared to the case of insulators, the experimental study of which is proving challenging.

The possibility of the nonlinear Hall effect has been reported in two ferrotoroidic candidate metals UNi$_4$B \cite{Oyamada18} and CuMnAs \cite{Godinho18}. However, in the former case, the contact resistance is too high to suppress a local Joule heating, and hence Seebeck effect may reflect the ghost of a giant nonlinear Hall effect even at the disordered state \cite{Mentink95_JMMM, Mentink95_PhysicaB}; this would also make it difficult to compare with the existing data on current-induced magnetization including its temperature dependence and amplitude \cite{Saito18}. By contrast, the latter one has already succeeded in observing and controlling the nonlinear Hall effect by deflecting N\'eel vector; this nonlinear response has been considered to come from the anisotropic magnetoresistance. Nevertheless, according to subsequent theoretical works, the N\'eel vector reflects the magnetic toroidal moment that induces a nonlinear Hall effect through the current-induced magnetization \cite{Hayami14, Watanabe18, Wang21}, although its direct experimental evidence is currently lacking.

%%%%%%%%%%%%%%%%%%%%%%%%%%%%%%%%%%%%
\begin{figure}[t]
\begin{center}
\includegraphics[width=1\linewidth]{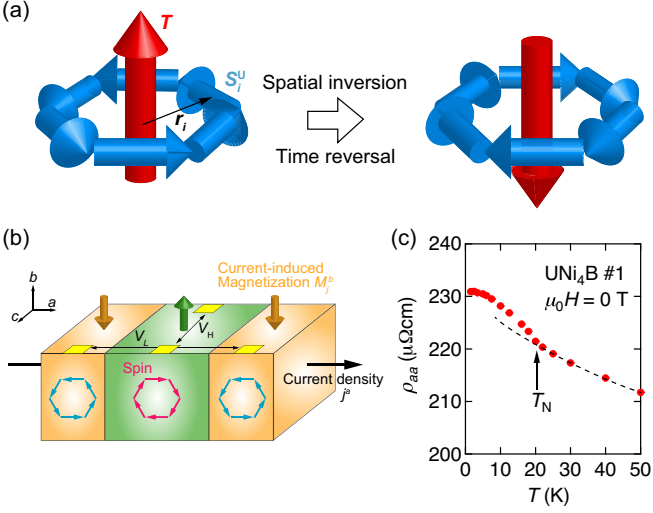}
\end{center}
\vspace{-6mm}
\caption{(color online). (a) Spatial-inversion and time-reversal symmetry in MT dipole, $\bm T$ (red arrow). As shown in Eq. (\ref{eq3}), this vector is defined as the outer product of the position from the inversion center (black arrow), ${\bm r}_i$, and the localized U moment (blue arrows), ${\bm S}^{\rm U}_i$. A spatial inversion reverses ${\bm r}_i$, but preserves ${\bm S}^{\rm U}_i$. By contrast, time reversal switches ${\bm S}^{\rm U}_i$ without any change in ${\bm r}_i$. Thus, $\bm T$ is odd under both symmetries. (b) Schematic illustration of the experimental setup. $\bm T$ consists of a spin vortex (pink/blue circle arrows). In the present work, $\bm T$ were aligned along the $c$ axis and a current density was applied along the $a$ axis (black arrow), $j^a$; this arrangement induces a magnetization, $\bm M_{j}$, along the $b$ axis (orange/green arrows) \cite{Saito18}. Its direction is affected by the MT domain expected in UNi$_4$B, as drawn by green and orange regions. We attached four contacts for measuring a longitudinal voltage ($V_L \parallel a$) and a Hall voltage ($V_{\rm H} \parallel c$). (c) Temperature dependence of zero-field longitudinal resistivity along the $a$ axis in sample \#1 of UNi$_4$B. The dashed line is an eye guide.} 
\vspace{-3mm}
\label{Fig1}
\end{figure}
%%%%%%%%%%%%%%%%%%%%%%%%%%%%%%%%%%%%

In this letter, to investigate the details of the nonlinear Hall effect for ferrotoroidic metals, we demonstrate carefully precise zero-field Hall measurements in UNi$_4$B with $T_{\rm N}\sim$ 20 K (namely, the ferrotoroidic transition temperature) for a wide temperature range from 1.5 K to 50 K as a model case. Below $T_{\rm N}$, a Hall voltage proportional to the square of current density is clearly observed even at zero field, whereas above $T_{\rm N}$ it is almost absent. This result consistently agrees with the theoretical expectation that the nonlinear Hall effect would originate from the current-induced magnetization, $\bm {M_j}$, given by the cross product of current density, $\bm j$, and MT dipole, $\bm T$:
\begin{equation}
\label{eq1}
\bm{V_{\rm H}} \sim \bm {M_j\times j} \sim \bm |{\bm j}|^2  \bm{T}. 
\end{equation}
Moreover, we have found that the current-induced magnetization estimated from our results through the above relation qualitatively and quantitatively scales with that actually measured in the previous research \cite{Saito18}. Our findings strongly indicate that the observed nonlinear Hall effect originates from the current-induced magnetization associated with the ferrotoroidal moments.

UNi$_4$B crystallizes in a slightly distorted hexagonal structure of U atoms \cite{Haga08, Takeuchi20, Tabata21}, but for simplicity, we treat it as a hexagonal structure in this text. Below $T_{\rm N}$, whereas the other 1/3 still remain in the paramagnetic state, 2/3 of the U atoms are ordered and form a vortex-like structure \cite{Mentink94, Mentink98, Willwater21, Mentink95_PRB, Lacroix96, Tejima97}; this can be regarded as the ferrotoroidal ordered state with MT dipoles along the $c$ axis \cite{Hayami15}, in which a net $b$-axis magnetization, $M_j^b$, has been reported to be generated by an applied current along the $a$ axis, $j^a$, as a magnetoelectric phenomenon (Fig.\ 1(b)) \cite{Saito18}. Although the $a$-axis longitudinal resistivity, $\rho_{aa}$, increases with decreasing temperature in the wide temperature range, its amplitude of $\sim$230 \textmu$\Omega$cm at low temperatures indicates a metallic state (Fig.\ 1(c)). Thus, UNi$_4$B is a good candidate for the study of the nonlinear Hall effect associated with ferrotoroidic metals.

Single crystals of UNi$_4$B were grown using the Czochralski method. The ingots were oriented by using the Laue photograph and then cut into a rectangular shape with the size of $\sim$0.96$\times$0.66$\times$0.08 mm$^3$ by using a spark cutter. The quality of the rectangular-shaped crystals was checked by the single crystal X-ray analysis and the magnetic susceptibility measurements. In the present work for the non-linear Hall effect, as shown in Fig.\ 1(b), an external current was applied along the $a$ axis and a Hall voltage was measured along the $c$ axis because the nonlinear Hall effect has been expected to follow Eq.\ (\ref{eq1}). To complementarily evaluate the nonlinear Hall effect, we used the combination of a direct current (DC) density, $j_{\rm dc}$, and an alternating current (AC) density, $j_{\rm ac}\sin(\omega t)$, with the frequency of $\omega/2\pi=$ 17 Hz. In this case, a Hall voltage is expected to follow
\begin{align}
\label{eq2}
V_{\rm H}&\equiv V_{\rm H}^{2\omega}\sin\left(2\omega t- \frac{\pi}{2}\right)+V_{\rm H}^{\omega}\sin(\omega t)+V_{\rm H}^{\rm 0} \notag \\
&=\alpha \left[\frac{j_{\rm ac}^2}{2} \sin\left(2\omega t- \frac{\pi}{2}\right) + 2j_{\rm dc}j_{\rm ac}\sin(\omega t) 
 + \frac{2j_{\rm dc}^2+j_{\rm ac}^2}{2} \right],
\end{align}
where $\alpha$ is a coefficient reflecting the size and direction of ${\bm T}$. Both of 2$\omega$ and $\omega$ components for $V_{\rm H}$ were evaluated as $V_{\rm H}^{2\omega}$ and  $V_{\rm H}^{\omega}$, respectively. It should be noted that the precise measurement of the DC component of $V_{\rm H}^{\rm 0}$ is very difficult due to the contact misalignment and the thermal drift. The improvement of the ratio of signal to noise was achieved by using both a cryogenic transformer and spot welding, obtaining a very low-noise voltage of $\sim$150 pV (for comparison of the noise with/without the cryogenic transformer, see Fig.\ S1 and Sec.\ S1 in Ref. \cite{SM}). Throughout this research, the variation of $V_{\rm H}^{2\omega}$ and $V_{\rm H}^{\omega}$ from those measured at 30 K, defined as $\Delta V_{\rm H}^{2\omega}$ and $\Delta V_{\rm H}^{\omega}$, respectively, were estimated to remove the background of a small but finite nonlinear signal distortion associated with a preamplifier and a function generator (for details, see Fig.\ S2 and Sec.\ S2 in Ref. \cite{SM}). Here, we stress that our conclusion does not change even if we use the values of $V_{\rm H}^{2\omega}$ and $V_{\rm H}^{\omega}$ at other temperatures above $T_{\rm N}$ as the background signal for the $2\omega$ and $\omega$ Hall components, respectively.

%%%%%%%%%%%%%%%%%%%%%%%%%%%%%%%%%%
\begin{figure*}[t]
\begin{center}
\includegraphics[width=0.85\linewidth]{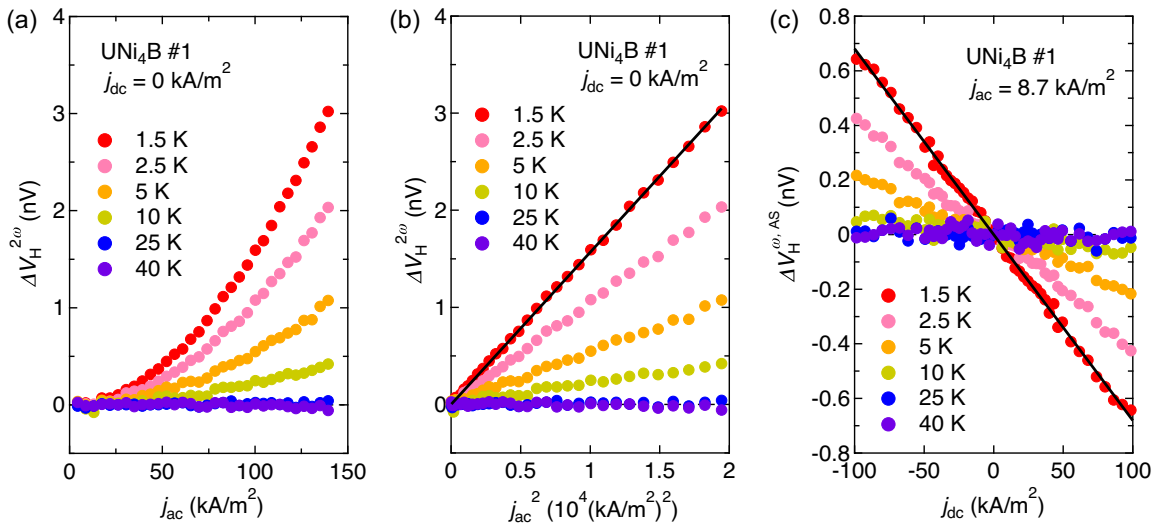}
\end{center}
\vspace{-6mm}
\caption{(color online). (a), (b) Temperature dependence of the 2$\omega$ component of Hall voltage in sample \#1 of UNi$_4$B. $\Delta V_{\rm H}^{2\omega}$ is plotted against (a) $j_{\rm ac}$ and (b) $j_{\rm ac}^2$. (c) $j_{\rm dc}$ dependence of $\Delta V_{\rm H}^{\omega, {\rm AS}}$ in sample \#1 of UNi$_4$B at the same temperatures as the $2\omega$ experiments shown in (a) and (b). The solid straight lines in (b) and (c) represent the line fittings for the collected data at 1.5 K, which gives the value of $\alpha$ from Eq. (\ref{eq2}).}
\vspace{-3mm}
\label{Fig2}
\end{figure*}
%%%%%%%%%%%%%%%%%%%%%%%%%%%%%%%%%%

To study the nonlinear Hall effect in UNi$_4$B, we first focus on the 2$\omega$ component in Eq. (\ref{eq2}). Figure\ 2(a) shows the $j_{\rm ac}$ dependence of $\Delta V_{\rm H}^{2\omega}$ at various temperatures crossing $T_{\rm N}$ for UNi$_4$B. Below $T_{\rm N}$, a nonlinear $j_{\rm ac}$ dependence of $\Delta V_{\rm H}^{2\omega}$ is clearly observed. This is essentially different from an ordinary Hall voltage proportional to an applied current; instead, this behavior follows a quadratic $j_{\rm ac}$ dependence expected for Eq.(\ref{eq2}), as shown in Fig.\ 2(b). The value of $\Delta V_{\rm H}^{2\omega}$ at $j_{\rm ac}\sim 140\ {\rm kA/m^2}$ is dramatically enhanced with reducing temperature and finally reaches $\sim$3 nV at 1.5 K. This enhancement is roughly consistent with the growth of ferrotoroidal moment consisting of magnetic dipole moments \cite{Mentink94, Mentink98, Willwater21, Nieuwenhuys95, Liu97}. Meanwhile, $\Delta V_{\rm H}^{2\omega}$ is virtually independent of both temperature and $j_{\rm ac}$ above $T_{\rm N}$; its value becomes almost zero even at $j_{\rm ac}\sim 140\ {\rm kA/m^2}$. This contrast signature above and below $T_{\rm N}$ implies the presence of nonlinear Hall effect only in the ferrotoroidic ordered state of UNi$_4$B.

As a complementary check of the nonlinear Hall effect in UNi$_4$B, we next evaluate the $\omega$ component in Eq. (\ref{eq2}). Figure\ 2(c) shows the antisymmetric part of $\Delta V_{\rm H}^{\omega}$, namely, $\Delta V_{\rm H}^{\omega, {\rm AS}}$, against $j_{\rm dc}$ for UNi$_4$B at the same temperatures as the 2$\omega$ experiments shown in Figs.\ 2(a) and (b). Here, we fixed $j_{\rm ac}$ as 8.7 kA/m$^2$, which is small enough compared to the controllable current range of DC bias. Below $T_{\rm N}$, a finite slope of $\Delta V_{\rm H}^{\omega, {\rm AS}}$--$j_{\rm dc}$ curve is clearly observed; this slope is further increased with decreasing temperature. In sharp contrast, above $T_{\rm N}$, its slope is approximately zero within experimental errors and does not change with temperature. This trend is well reproduced in a different sample (Fig.\ S3 and Sec.\ S3 in Ref. \cite{SM}). In the $\omega$ measurements, either of $j_{\rm dc}$/$j_{\rm ac}$ is utilized to induce the net magnetization and the opposite one is used to measure the Hall voltage, whereas in the $2\omega$ measurements, $j_{\rm ac}$ simultaneously plays both roles. Regardless of the different measurement methods, the results of the $2\omega$ and $\omega$ components are in qualitative agreement with each other. We now try to quantitatively compare the 2$\omega$ and $\omega$ components by calculating the value of $\alpha$ in Eq.\ (\ref{eq2}). Figure 3 shows the temperature dependence of $\alpha$ estimated from the 2$\omega$ (blue square) and $\omega$ (red circle) components. Clearly, the sign and value of $\alpha$ in both components are approximately scaled with each other for all measured temperatures, including its dramatic enhancement below $T_{\rm N}$ and its temperature-independent tiny value above $T_{\rm N}$. This predicted coincidence provides unambiguous evidence for the zero-field current-induced Hall effect in the ferrotoroidic metal UNi$_4$B.

An important question arising here is what is the origin of the observed current-induced Hall effect in UNi$_4$B. According to previous theoretical and experimental studies, $M_j^b$ is generated by applying $j^a$ through the ferrotoroidal moment parallel to the $c$ axis \cite{Hayami14, Saito18}; this immediately implies that the zero-field Hall effect would originate from the current-induced magnetization. To experimentally confirm this point, we here estimate the value and temperature dependence of $M_j^b$ from our Hall effect results. Assuming the linear $j^a$ dependence of $M_j^b$, the relation of $M_j^b=\alpha\frac{\chi^b}{R_{\rm H}^{\rm 0}w}j^a$ was derived from the measurements of magnetization and ordinary Hall effect in the zero-field limit (for details, see Fig.\ S4 and Sec.\ S4 in Ref. \cite{SM}). Here, $w$ is the width of the sample, and $\chi^b$ and $R_{\rm H}^{\rm 0}$ represent the magnetic susceptibility and the ordinary Hall coefficient in the zero-field limit of ${\bm H}\parallel b$, respectively. From the observed $\alpha$, one finds that at $j^a\sim55.6$ kA/m$^2$, the estimated current-induced magnetization exhibits a similar behavior as the results in previous direct measurements \cite{Saito18} (see Fig.\ 4); their temperature dependencies are scaled appropriately, and their magnitudes are of the same order. The magnitude difference is probably because of the presence of MT domain shown in Fig.\ 1(b) (see also Sec.\ S3 in Ref. \cite{SM}). Moreover, surprisingly, we confirm that the values of current-induced magnetization and Hall conductivity roughly follow the linear relation of spontaneous magnetization and anomalous Hall conductivity for the ordinary ferromagnets (for details, see Fig.\ S5 and Sec.\ S5 in Ref. \cite{SM}). These comparisons support the conclusion that the current-induced magnetization is most likely to drive the nonlinear `anomalous'  Hall effect in UNi$_4$B at zero field.

%%%%%%%%%%%%%%%%%%%%%%%%%%%%%%%%%%
\begin{figure}[t]
\begin{center}
\includegraphics[width=0.75\linewidth]{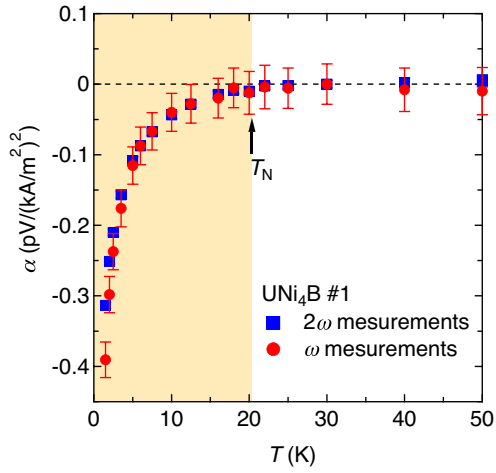}
\end{center}
\vspace{-7mm}
\caption{(color online). Temperature dependence of $\alpha$ for the $2\omega$ (blue squares) and $\omega$ (red circles) components in sample \#1 of UNi$_4$B. The shaded region represents the ferrotoroidic ordered phase.} 
\vspace{-3mm}
\label{Fig3}
\end{figure}
%%%%%%%%%%%%%%%%%%%%%%%%%%%%%%%%%%

It is noteworthy that two mechanisms can trigger the current-induced magnetization in noncentrosymmetric metals: magnetoelectric effect \cite{Curie94, Hayami14} and Edelstein effect \cite{Edelstein90}. The former is active in the absence of both spatial-inversion and time-reversal symmetry, regardless of metals or insulators; this may apply to the present case, in which the temperature evolution of $M_j^b$ is expected to follow that of $|{\bm T}|$, as seen in the forms of Eq.\ (\ref{eq1}). Nevertheless, all of neutron diffraction, muon spin rotation ($\mu$SR), and nuclear magnetic resonance (NMR) measurements have pointed out that the temperature dependence of the localized U moment ${\bm S}^{\rm U}_i$ is clearly different from the case of $M_j^b$. This is curious because both of ${\bm S}^{\rm U}_i$ and $M_j^b$ should give a similar temperature dependence from the viewpoint of its definition; $|{\bm T}|$ is defined as the summation of the vector products of ${\bm S}^{\rm U}_i$ and a virtually temperature independent position vector ${\bm r}_i$ for magnetic sites  $i$, namely
\begin{align}
\label{eq3}
{\bm T}=\frac{g\mu_{\rm B}}{2}\sum_i{\bm r}_i{\bm \times}{\bm S}^{\rm U}_i,
\end{align}
where $g$ is the gyromagnetic ratio, $\mu_{\rm B}$ is the Bohr magneton, and the summation is taken over an appropriate magnetic basis. This discrepancy likely stems from the band effect in metals, namely that the growth of $|{\bm T}|$ leads to the change in the electronic state close to Fermi level. The latter is another scenario because a magnetic structure without parity-time inversion symmetry still remains possible in the case of UNi$_4$B \cite{Saito18}. In this circumstance, an applied electric current causes the different redistribution of conduction band with opposite spin textures, which can result in the temperature dependence of the net spin polarization irrespective of $|{\bm T}|$.

%%%%%%%%%%%%%%%%%%%%%%%%%%%%%%%%%%
\begin{figure}[t]
\begin{center}
\includegraphics[width=0.79\linewidth]{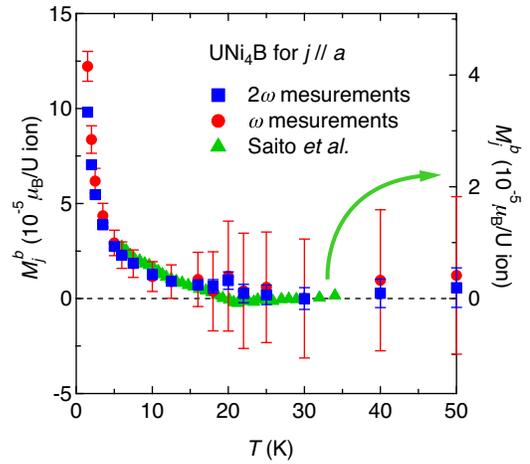}
\end{center}
\vspace{-6mm}
\caption{(color online). Temperature dependence of current-induced magnetization $M_j^b$ at $j^a\sim 55.6$ kA/m$^2$ for UNi$_4$B. In the left axis, the blue squares and red circles represent $M_j^b$ estimated from the $2\omega$ and $\omega$ measurements in sample \#1 of UNi$_4$B, respectively. In the right axis, the green triangles show the directly measured value of $M_j^b$ \cite{Saito18}.} 
\vspace{-3mm}
\label{Fig4}
\end{figure}
%%%%%%%%%%%%%%%%%%%%%%%%%%%%%%%%%%

Besides the above two possibilities, the anomaly at $T^*\sim0.33$ K of unknown origin in measurements of resistivity and specific-heat for UNi$_4$B \cite{Movshovich99, Mydosh99} may explain the above-mentioned inconsistency. Very recently, it has been suggested that this anomaly does not reflect the additional modification of magnetic structure \cite{Willwater21}, but the contribution of electric-quadrupoles of the paramagnetic 1/3 U ions \cite{Oyamada07, Kishimoto18, Yanagisawa21}. The quadrupolar contributions in the ferrotoroidic ordered state seem to affect the drastic enhancement of current-induced magnetization with decreasing temperature, especially below $\sim5$ K where the sudden softning of transverse ultrasonic mode C$_{66}$ is observed \cite{Yanagisawa21} in addition to the reduction of $\chi^b$ \cite{Saito18} and $\rho_{aa}$ \cite{Mentink94} as well as the gradual enhancement of specific heat divided by temperature \cite{Mentink94} and thermal expansion along the $c$ axis \cite{Mentink97}. Further theoretical studies are required to elucidate the relation between $\alpha$ and $|{\bm T}|$.

In summary, to examine the origin of zero-magnetic-field current-induced Hall effect in the ferrotoroidic candidate material UNi$_4$B, we have performed precise Hall effect measurements in the wide temperature range covering $T_{\rm N}$ by using a cryogenic transformer with very low noise. From the measurements of the $2\omega$ and $\omega$ components for the Hall voltage, we have found that a finite nonlinear Hall voltage gradually grows below $T_{\rm N}$. The current-induced magnetization estimated from our Hall signals is quite in agreement with the previous directly measured value. These results demonstrate the presence of nonlinear anomalous Hall effect that originates from the current-induced magnetization through the ferrotoroidal moments.

%%%%%%%%%Acknowledgments
We thank H. Amitsuka, K. Hattori, S. Hayami, H. Hidaka, T. Ishitobi, H. Kusunose, K. Matsuura, Y. Motome, C. Tabata, H. Watanabe, Y. Yanase, and M. Yatsushiro for helpful discussions. This work was supported by Grants-in-Aid for Scientific Research (Grants Nos. 17H02920, 18H01167, 18K18730, 19H00646, 20K20889, and 20K20901) from MEXT and JSPS, by a Grant-in-Aid for Scientific Research on Innovative Areas "J-Physics: Physics of Conductive Multipole Systems" (KAKENHI Grant No. JP15H05884), "Topological Materials Science" (KAKENHI Grant No. JP18H04213), and "Quantum Liquid Crystals" (KAKENHI Grant No. JP20H05162) from JSPS of Japan, and by Multidisciplinary Research Laboratory System (MRL), Osaka University.

\end{document}